\documentclass[journal=ancac3,manuscript=article]{achemso}

\usepackage{graphicx}
\usepackage{dcolumn}
\usepackage{bm}
\usepackage{hyperref}
\usepackage{xcolor}
\usepackage{amssymb}
\usepackage{amsfonts}
\usepackage{amsmath}
\usepackage{graphicx}
\usepackage{mwe}
\usepackage{ulem}%

\title{Optically induced spin electromotive force 
	in ferromagnetic-semiconductor quantum well structure} 

\author{
	Igor~V.~Rozhansky}
\affiliation{Ioffe Institute, St.~Petersburg, 194021, Russia}
\email{rozhansky@gmail.com}

\author{
	Ina~V.~Kalitukha}
\affiliation{Ioffe Institute, St.~Petersburg, 194021, Russia}
\author{
	Grigorii~S.~Dimitriev}
\affiliation{Ioffe Institute, St.~Petersburg, 194021, Russia}
\author{
	Olga~S.~Ken}
\affiliation{Ioffe Institute, St.~Petersburg, 194021, Russia}
\author{
	Mikhail~V.~Dorokhin}
\affiliation{Lobachevsky State University of Nizhny Novgorod, Nizhny Novgorod, 603950, Russia}
\author{
	Boris~N.~Zvonkov}
\affiliation{Lobachevsky State University of Nizhny Novgorod, Nizhny Novgorod, 603950, Russia}
\author{
	Dmitri~S.~Arteev}
\affiliation{Ioffe Institute, St.~Petersburg, 194021, Russia}
\author{
	Nikita~S.~Averkiev} 
\affiliation{Ioffe Institute, St.~Petersburg, 194021, Russia}
\author{
	Vladimir~L.~Korenev}
\affiliation{Ioffe Institute, St.~Petersburg, 194021, Russia}


\begin{document}
	
	\begin{abstract}
			Hybrid structures combining ferromagnetic (FM) and semiconductor constituents have great potential for future applications in the field of spintronics. 
			A systematic approach to study spin-dependent transport in the GaMnAs/GaAs/InGaAs quantum well (QW) hybrid structure with a few nanometer thick GaAs barrier is developed.  
			It is demonstrated that a  combination of spin electromotive force measurements and photololuminescence detection provides a powerful tool for studying the properties of such hybrid structures and allows to resolve the dynamic FM  proximity effect on a nanometer scale. The method can be generalized on various systems including rapidly developing 2D van der Waals materials.
	\end{abstract}
	
	Today, the integration of magnetism into semiconductor (SC)
	electronics is a relevant task that is to be solved yet.
	Magnetic materials are capable of non-volatile storing large amounts of data. Hybrid ferromagnet-semiconductor quantum well (FM-QW) systems are potentially an ideal platform for such integration. The fundamental interest in such systems is associated with the presence of new spin-spin interactions, which exist upon bringing the FM and SC into a contact. 
	In this case the FM proximity effect arises, i.e. spin 
	polarization of the charge carriers in a semiconductor in a vicinity of a ferromagnet.
	A long-range interaction between the spins of QW holes and FM magnetic atoms has been found in hybrid structures based on Co,Fe/CdTe QW~\cite{Korenev2016,Korenev2019}. 
	
	A nontrivial phenomenon of carriers spin polarization was also found in the GaAs/InGaAs system with a GaMnAs FM layer adjacent to the InGaAs QW. In the absence of optical excitation, the spin polarization of the majority charge carriers (holes) in the QW was detected electrically~\cite{Pankov2009}
	and was explained by the short-range exchange interaction of holes in the QW with the FM. It turned out, however, that under optical excitation the situation changes radically: the proximity effect is dynamic. The polarization kinetics 
	demonstrates the spin-dependent transport of photoelectrons through the GaAs barrier~\cite{Korenev2012,Akimovpss,Myers2004,Zaitsev2010}.
	In the case of charge transfer, there should also be a spin-dependent contribution to the photo-electromotive force, which we will call the spin-photo-EMF. Naturally, under the conditions of charge and spin transfer through the interface, the magnetic hysteresis loops of the spin photo-voltage and PL circular polarization in the hybrid FM-QW system should correlate with each other. Such a study has not been carried out to date.

Spin-dependent transfer is manifested in three  spectacular effects: (i) PL circular polarization under unpolarized excitation, (ii) dependence of the PL intensity from the QW on the circular polarization degree of the excitation, and (iii) spin-dependent photo-voltage across the junction. Two conditions should be fulfilled to unambiguously isolate the tunneling spin transfer. First is the choice of the FM where the magnetic circular dichroism (MCD) is absent. The second is resonant excitation of the QW, otherwise the above-barrier drift-diffusion spin transport over macroscopic distances $\sim 10$ $\mu$m hinders the FM-induced proximity effect~\cite{Endres2013,PhysRevB.64.121201}. In our case QW embedded into the structure near the interface sets the nanometer scale depth resolution of proximity magnetic effect, provided that the QW is excited resonantly~\cite{Korenev2019}.
	
For hybrid structures with a few-nanometer-thick barrier, PL signal under resonant excitation is too weak due to the strong non-radiative recombination of the photoexcited carriers, which tunnel from the QW into the FM. On the contrary, the spin-photo-voltage signal in such case is reliably measured and thus appears to be a working tool to study the magnetic proximity effect on a nanometer spatial scale, where it is  inaccessible for resonant-PL technique. This work comes up with an idea of a complex analysis of the magnetic proximity effect by combining the optical and electrical detection of the spin-dependent electron transfer with nanoscale spatial resolution. We study  GaMnAs/GaAs/InGaAs heterostructure upon resonant optical excitation of the InGaAs QW located at a distance of only 5–10 nm from the nanometer-thick FM \mbox{GaMnAs} layer. A unified theoretical model is presented capable of describing the entire set of experiments. We revealed the interplay between the tunneling rate and the recombination rate of the carriers in the QW. This finding is crucial to the problem of spin injection~\cite{PhysRevB.62.R16267,PhysRevB.67.201304}, well-known resistance mismatch issue~\cite{PhysRevB.62.R4790} and recently reported times mismatch effect~\cite{Safarov2022}. 
	
Our complementary approach presents a powerful tool to identify spin-dependent transport in the FM-QW hybrid structures and can also be generalized on various systems including rapidly developing 2D  van der Waals materials~\cite{ZUTIC201985}.	
	
\begin{figure}[h]
		\centering
		\includegraphics[width=0.5\textwidth]{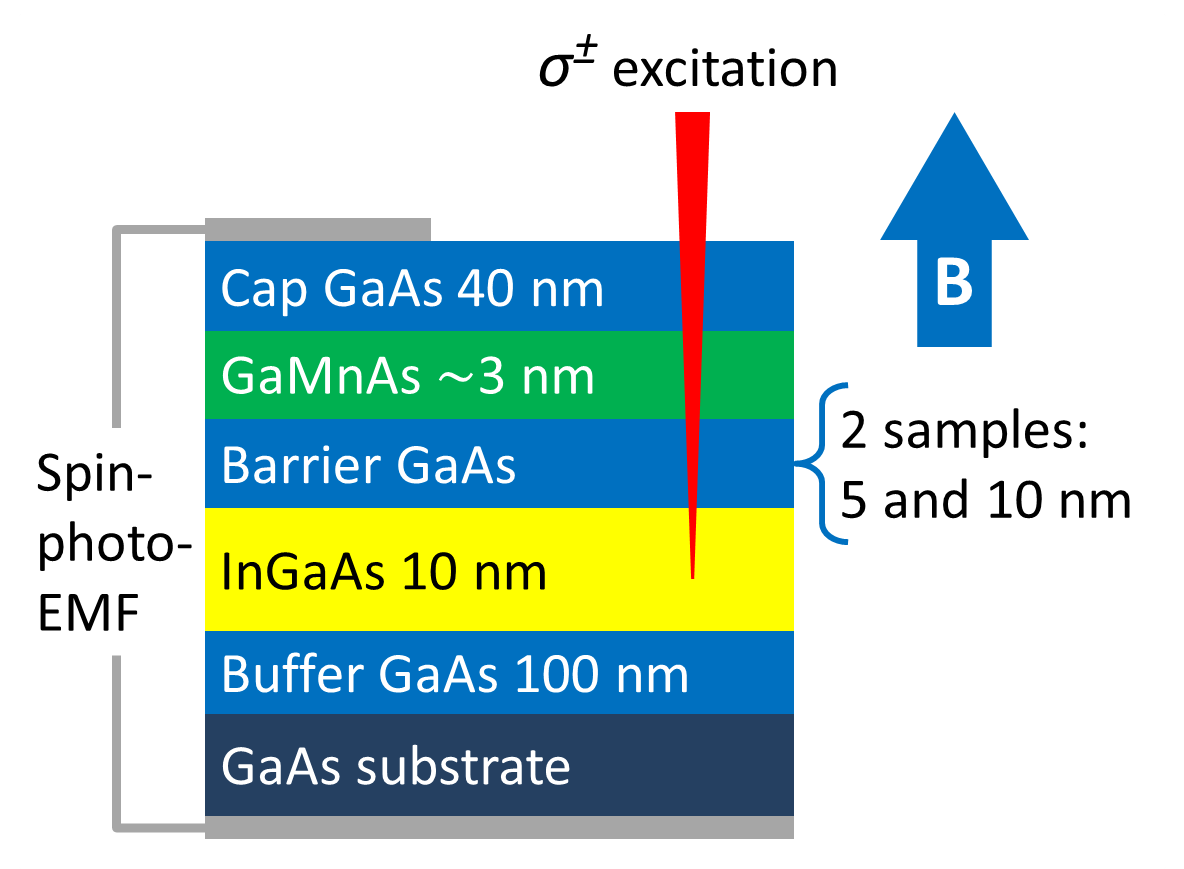}
		\caption{Schematic representation of the GaAs/GaMnAs/GaAs/InGaAs QW hybrid structure showing also the scheme of the spin-photo-EMF experiment.}
		\label{sample}
	\end{figure}

	\textbf{Experiment.} In this paper we study a GaMnAs/GaAs/In$_x$Ga$_{1-x}$As QW hybrid structure with In content $x\sim10\%$ (Fig.~\ref{sample}). 10~nm InGaAs QW  and the FM GaMnAs layer (thickness of about $3$~nm) are separated by a nonmagnetic GaAs barrier. We study structures with barrier thicknesses $d=5$~nm and $d=10$~nm, marked as "$5$~nm sample" and "$10$~nm sample", correspondingly. Contacts (grey layers in Fig.~\ref{sample}) are attached to the front surface of the samples (strongly doped p-type due to the formation of the GaMnAs) and to the n-GaAs substrate. More information on the sample design can be found in the Supplementary Information (SI), section S1.
	
	Measurements of the photo-voltage in the open circuit mode gives the value of the photo-induced EMF. To measure the spin-independent photo-induced EMF, which we will call photo-EMF, the sample is excited with linearly ($\pi$) polarized light of modulated intensity. To measure the spin-dependent part of the photo-induced EMF (spin-photo-EMF), the sample is excited with alternating $\sigma^+/\sigma^-$ circularly polarized light of constant intensity. The amplitudes of the modulated photo-EMF signal $U_I$ and spin-photo-EMF signal $U_S$ are measured using the lock-in amplifier. The sketch of the spin-photo-EMF experiment is demonstrated in Fig.~\ref{sample}. For more details on the electrical and optical measurements see SI (S1).
	
	In the experimental section, we chose to present the results for the $10$~nm sample, as full and comprehensive, while the results for the $5$~nm sample can be found in the SI, S3.
	
\begin{figure}[h]
		\centering
		\includegraphics[width=0.75\textwidth]{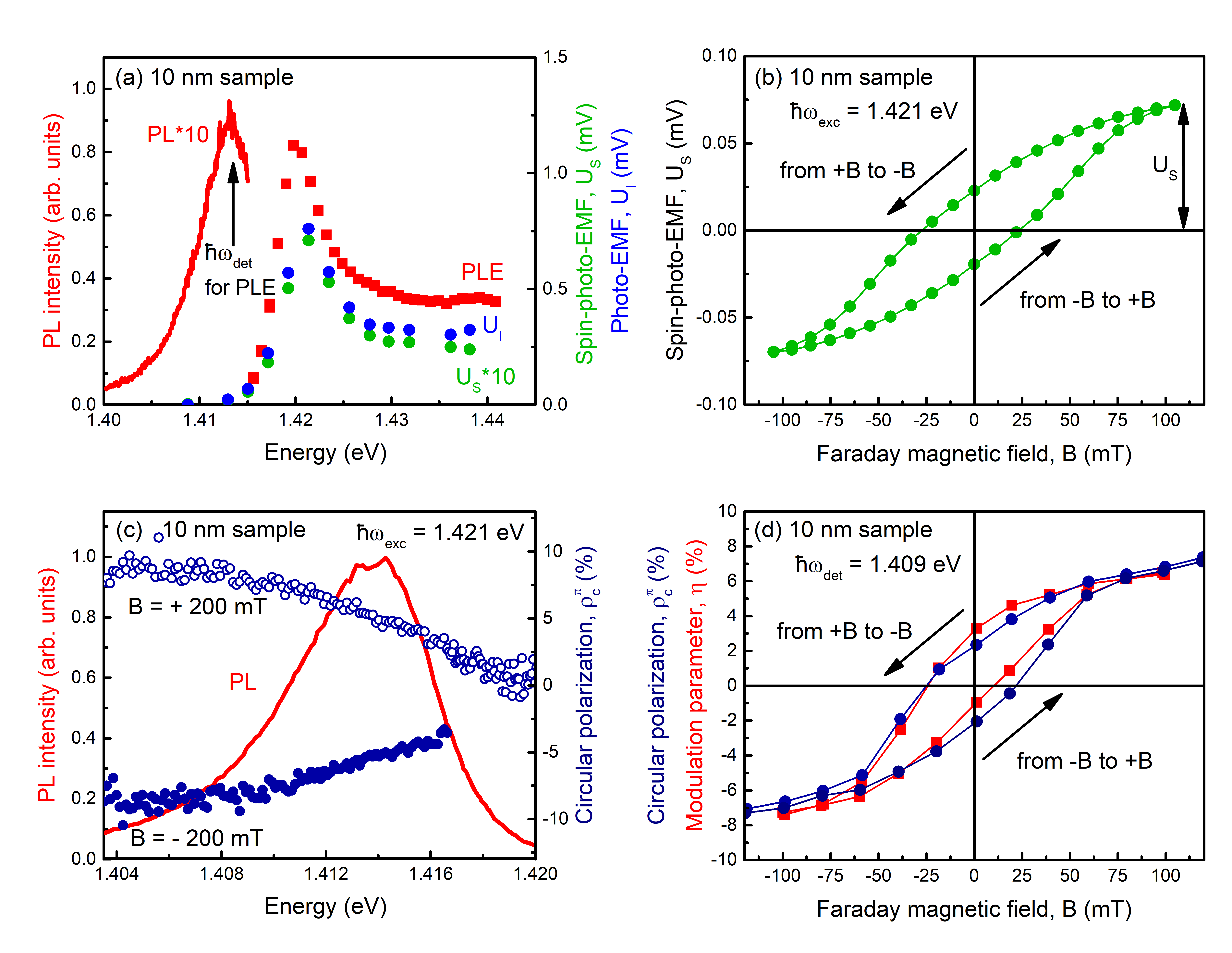}
		\caption{$10$~nm sample (a) PL spectrum of the QW (solid red line), PLE spectrum with detection energy $1.4135$~eV (red squares) on the left scale. Spectra of the photo-EMF $U_I(\hbar \omega_{exc})$ (blue circles) and the spin-photo-EMF $U_S(\hbar \omega_{exc})$ (green circles, multiplied by 10) on the right scale. The spin-photo-EMF is measured in magnetic field $B=100$~mT applied in Faraday geometry. (b) Hysteresis loop of the spin-photo-EMF $U_S(B)$ is shown for magnetic field sweep from $B=-100$~mT to $B=+100$~mT, and the opposite direction. (c) PL spectrum of the QW is shown with solid red line, circular polarization degree $\rho^{\rm \pi}_{\rm c}$ in Faraday magnetic field is shown with open circles ($B=+200$~mT) and closed circles ($B=-200$~mT). (d) Magnetic field dependencies $\rho^{\rm \pi}_{\rm c}(B)$ (blue circles) and $\eta(B)$ (red squares) are shown for magnetic field sweep from $B=-100$~mT to $B=+100$~mT, and the opposite direction. All the experiments are carried out at the temperature of $T=2$~K.}
		\label{emf1}
	\end{figure}

	Spectra of the photo-EMF $U_I(\hbar \omega_{exc})$ and spin-photo-EMF $U_S(\hbar \omega_{exc})$, where $\hbar \omega_{exc}$ is laser energy, are shown in Fig.~\ref{emf1}a. To magnetize the FM layer magnetic field $B=100$~mT is applied to the sample in Faraday geometry (magnetic field is parallel to the optical axis and structure growth axis, see Fig.~\ref{sample})~\cite{Akimovpss}. The photo-EMF spectrum reaches its maximum of $0.8$~mV at excitation energy $\hbar \omega_{exc}=1.421$~eV. The maximum of the spin-photo-EMF spectrum is approximately 10 times smaller and reaches $0.07$~mV under the same excitation energy. Maxima of the photo-EMF and spin-photo-EMF spectra coincides with the peak in PL excitation (PLE) spectrum (red squares in Fig.~\ref{emf1}a), while maximum in the PL spectrum (Fig.~\ref{emf1}a, red line) is shifted to the lower energies by $\sim 10$~meV. It is reasonable to associate the PLE maximum with absorption on the free exciton in the QW, while the PL maximum -- with recombination of the exciton localized in the plane of the QW.
	
	The amplitude of the photo-EMF does not depend on the magnetic field up to $150$~mT. Magnetic field dependence of the spin-photo-EMF $U_S(B)$ shows hysteresis loop with saturation in $B_{sat}=100$~mT (Fig.~\ref{emf1}b) and weakly depends on the laser energy. Such a behavior resembles the ferromagnet magnetization curve. This similarity points to spin-dependent tunneling of charge carriers from the QW into the FM layer as a mechanism of the spin-photo-EMF. It is worth noting that the contribution of the MCD is negligible ($< 0.1\%$), see SI, S2.
	To quantify the spin-dependent contribution to the photo-EMF, we introduce the relative spin-photo-EMF $\xi$ independent of the absorption coefficient:
	
	\[
	\xi = \frac{U^+(g,B)-U^-(g,B)}{U^+(g,B)+U^-(g,B)} =
	\frac{U_S}{U_I},
	\]
	where $U^\pm(g,B)$ are photo-EMF values in magnetic field $B$, under laser excitation with power density $g$ and circular polarization $\sigma^+$ and $\sigma^-$ correspondingly. For more details see SI (S1). In the $10$~nm sample $\xi \approx 10\%$ and weakly depends on the excitation energy in the range of $1.42 - 1.44$~eV.
	
	A mechanism of the spin-photo-EMF formation is the spin-dependent tunneling of the charge carriers from the QW into the FM layer. This conclusion is consistent with the optical measurements of the dynamic proximity effect - interaction of the spin system of the charge carriers in the QW with that of the FM~\cite{Korenev2012}. Two main parameters were measured: 1) Degree of the PL circular polarization under linear ($\pi$) excitation $\rho^{\rm \pi}_{\rm c} = (I^{\pi}_{\sigma+}-I^{\pi}_{\sigma-})/(I^{\pi}_{\sigma+}+I^{\pi}_{\sigma-})$, where $I^{\pi}_{\sigma+}$ and $I^{\pi}_{\sigma-}$ are the intensities of the $\sigma^+$ and $\sigma^-$ polarized PL components under $\pi$ excitation, respectively, and 2) PL intensity modulation parameter $\eta$ measured as total PL intensity under modulated circular polarization $\sigma^+$/$\sigma^-$ of the laser excitation $\eta=(I^{\sigma+}-I^{\sigma-})/(I^{\sigma+}+I^{\sigma-})$. Figure~\ref{emf1}c shows PL spectrum of InGaAs QW (red line) with maximum at $1.414$~eV (exciton recombination) and the degree of the PL circular polarization $\rho^{\rm \pi}_{\rm c}$ in small magnetic field in Faraday geometry ($B=\pm 200$~mT). Figure~\ref{emf1}d shows magnetic field dependence of $\rho^{\rm \pi}_{\rm c}$ (blue circles) at the detection energy $\hbar \omega_{det}=1.409$~eV. The dependence $\rho^{\rm \pi}_{\rm c}(B)$ exhibits hysteresis with saturation around $B_{sat}=100$~mT. Magnetic field dependence of the intensity modulation parameter $\eta(B)$ is shown by red squares in Fig.~\ref{emf1}d and exhibits behaviour similar to that of $\rho^{\rm \pi}_{\rm c}(B)$. Presence of hysteresis indicates the interaction of charge carriers in the QW with the FM i.e. the FM proximity effect.
	
	The two measured quantities have the following physical meaning. The PL intensity modulation parameter is determined by the difference in the PL intensity under $\sigma^+$ and $\sigma^-$ circularly polarized excitation in Faraday magnetic field. For one excitation helicity, spin polarized electrons are generated with spin along the magnetization of the FM layer, and in the other case in the opposite direction. Due to the spin-dependent tunneling in one case more electrons are captured from the QW into the FM. This dynamic effect leads to unequal PL intensity from the QW under different signs of the circular polarization of excitation.
	
	In its turn, the PL circular polarization degree $\rho^{\rm \pi}_{\rm c}$ equals the degree of the spin polarization of electrons in the QW $\rho^{\rm \pi}_{\rm c}=-P_e$ (minus here reflects the selection rules). Holes in the QW are unpolarized so the main contribution comes from spin polarized electrons~\cite{Korenev2012}. In case of the spin-dependent tunnelling, spin polarization of electrons $P_e$ originates from different capture rates of electrons with spins along and against the magnetization direction in the FM, and $\rho^{\rm \pi}_{\rm c} \approx \eta$~\cite{Korenev2012}.
	
	Our experiment shows that dependencies $\rho^{\rm \pi}_{\rm c}(B)$ and $\eta(B)$ (Fig.~\ref{emf1}d) coincide and are in good agreement with the magnetic field behavior of the spin-photo-EMF $U_S(B)$ (Fig.~\ref{emf1}b). Thus, both optical and electrical experiments  are determined by the spin-dependent electron capture from the QW to the FM layer complementing each other.
	
	The same electrical and optical experiments are performed on sample with $5$~nm thick barrier (see SI, section S3). For this sample the value of the relative spin-photo-EMF $\xi = \frac{U_S}{U_I} \approx 0.5\%$ is 20 times less than for the sample with $10$~nm barrier, where $\xi \approx 10\%$, although the tunnelling through the barrier has increased. This surprising fact has a simple explanation. In the case of the thin barrier electrons are effectively captured into the FM independently of their spin polarization. This leads to almost complete charge separation, so that the photo-EMF does not depend on $\sigma^+$/$\sigma^-$ laser modulation.
	
	Main results obtained in experiment are summarized in Table~\ref{tab}. Note that for $5$~nm sample PL intensity under resonant excitation of the QW is negligibly small, which impedes measuring the intensity modulation parameter $\eta$. In this case the spin-photo-EMF turns out to be a more sensitive tool to indicate spin-dependent tunneling of charge carriers from the QW to the FM.
	
	\begin{table}[h!]
		
		\begin{tabular}{|p{1.5cm}|p{1.5cm}|p{1.5cm}|p{1.5cm}|}
			\hline
			$d$, nm & $\xi = \frac{U_S}{U_I}$ & $\rho^{\rm \pi}_{\rm c}(B_{sat})$ & $\eta(B_{sat})$  \\
			[1mm]\hline
			5 & 0.5$\%$ & 9$\%$ & - \\
			[1mm]\hline
			10 & 10$\%$ & 8$\%$ & 6$\%$\\
			[1mm]\hline
		\end{tabular}
		\caption{Experimental results for samples with barrier thickness $d=5$~nm and $d=10$~nm.}
		\label{tab}
	\end{table}
	
\begin{figure}[h]
		\centering
		\includegraphics[width=0.4\textwidth]{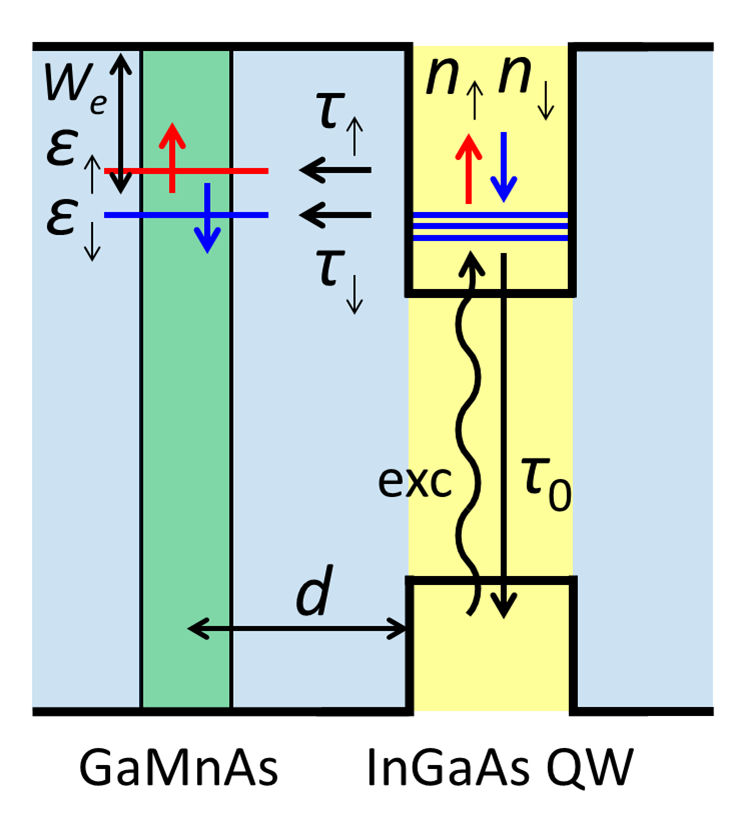}
		\caption{Scheme  of the hybrid FM-QW
			heterostructure.}
		\label{fig1}
	\end{figure}

	\textbf{The model.} The model considering spin dynamics is illustrated by Fig.~\ref{fig1}.
	The structure consists of  
	In${_x}$Ga$_{1-x}$As QW and a remote Mn-doped layer located at a few nanometers from the QW. 
	Along with the acceptor-like states formed by Mn substituting Ga, there are interstitial  Mn$_{I}$ double-donor states, which appear to be split in the 
	spin projection into the energy levels $\varepsilon_\downarrow,\varepsilon_\uparrow$ by the exchange field in the ferromagnetic Mn-doped layer~\cite{OURPRB2015}.
	Taking into account the donor level spin splitting we note that there are two tunneling channels corresponding to the opposite electron spin projections.
	We define by $\tau_{\uparrow},\tau_{\downarrow}$ the tunneling times (inverse tunneling rates) 
	of electron tunneling from the QW to the donor states, $\uparrow$, $\downarrow$ 
	correspond to the electron 
	spin projections
	on $z$ axis being the external magnetic field  direction. The difference $\Delta\tau=\tau_\uparrow-\tau_\downarrow$  is known to lead to dynamic spin polarization of the electrons in the QW~\cite{Korenev2012, OURPRB2015}. 
	Here we present a rate equations model allowing for calculation of spin-dependent EMF under open circuit conditions. 
	Let us assume the  light absorption in the QW leads to generation of the non-equilibrium electrons with a total generation rate $G_0$.
	Due to selection rules the circular polarized light results in  
	different generation rates $G^\pm_{\uparrow\downarrow}$
	for the electrons with opposite spin projections.  Here the upper index indicates the 
	sign of the light circular polarization and the lower index 
	describes the spin projection of the generated electrons. 
	Denoting the magnitude of the electron polarization degree due to optical orientation by $P_0$ we have  
	\begin{align}
		{G^+_ {\uparrow} } + {G^+_\downarrow } &=
		G^-_ {\downarrow}  + {G^-_\uparrow }=
		G_0 \nonumber\\ 
		{G^+_ {\uparrow} } - {G^+_\downarrow } &=
		G^-_ {\downarrow}  - {G^-_\uparrow }=-
		{P_0}G_0.\end{align}
	The minus sign in the right-hand part indicates that at $\sigma_+$ circular polarization there are more photogenerated spin-down electrons.    
	
	As MCD is negligible (see  SI), 
	the generation rate $G_0$ does not depend on the light helicity.    Denoting the concentration of the photogenerated electrons by 
	$n^\pm_{\downarrow\uparrow}$ we
	have the balance
	\begin{equation}{G^\pm_\uparrow } = \frac{{{n^\pm_\uparrow }}}{{{\tau _ \uparrow }}}
		+\frac{{{n^\pm_\uparrow }}}{{{\tau_0 }}}
		, \quad
		{G^\pm_\downarrow } = \frac{{{n^\pm_\downarrow }}}{{{\tau _ \downarrow }}}+\frac{{{n^\pm_\downarrow }}}{{{\tau_0 }}},
		\label{eq:bal2}
	\end{equation}
	where, along with the tunneling, the  spin-independent radiative recombination is accounted for by  the time $\tau_0$.  
	The photogenerated electrons form 
	spin-dependent current flowing into the bound states in the p-type region thus charging it negatively.  For the open circuit conditions the lowering of the potential barrier results in the opposite current of the  majority carriers that compensates the photocurrent. Hence one can apply the conventional expression for photo-EMF 
	appearing at the p-n junction: 
	\begin{equation}
		U^{\pm}=\frac{kT}{e}\ln\left(1+\frac{I^\pm_{ph}}{I_s}\right)\approx \frac{2\alpha}{e} I^\pm_{ph},\quad \alpha=\frac{kT}{ 2 I_s}
		\label{eq:emf}
	\end{equation}
	here $I^\pm_{ph}$ is the photocurrent density under the right (+) or left (-) circular polarization of the optical pumping  and 
	$I_s$ is the saturation current density of the p-n junction, 
	$U^{\pm}$ is the corresponding EMF,  
	$e$ is the electron charge, $T$ is the temperature, $k$ is the Boltzmann constant; in the linear regime $I^\pm_{ph}\ll I_s$.
	The photocurrent of the carriers with a given spin
	depends on their density and the tunneling rate, 
	summing for both spin channels we write: 
	\begin{equation}
		I^\pm_{ph}=e\left(\frac{n^\pm_\downarrow}{\tau_\downarrow}+\frac{n^\pm_\uparrow}{\tau_\uparrow}\right).
		\label{eq:iph}
	\end{equation}
	
	\textbf{\textit{Electrically detected FM proximity effect}}. 
	The photo-EMF for a right  and left  circular polarization of the incident light is   different. Using  (\ref{eq:iph}),  (\ref{eq:emf}) and (\ref{eq:bal2}) 
	we obtain spin-photo-EMF:
	\begin{equation}
		U_S = \frac{ U^ +  - U^ - }{2}=-\alpha  {G_0}{P_0}{\tau _0}\frac{{{\tau _ \downarrow } - {\tau _ \uparrow }}}{{\left( {{\tau _ \uparrow } + {\tau _0}} \right)\left( {{\tau _ \downarrow } + {\tau _0}} \right)}}.
	\end{equation}
	Note that a finite $\tau_0$ is essential otherwise 
	all the photogenerated carries eventually contribute to the EMF regardless their spin making $U^-=U^+$.
	Similar effect  has been noted in Ref.~\cite{Safarov2022}. 
	The tunneling time difference is due to the spin split of the bound states in the FM layer being linear
	to its magnetization $M$.
	Let us introduce 
	\begin{align}
		{\tau} = \frac{{{\tau _ \downarrow } + {\tau _ \uparrow }}}{2} \quad \Delta \tau  = {\tau _ \downarrow } - {\tau _ \uparrow }=\gamma\tau M.
	\end{align}
	The polarization-independent 
	contribution to the EMF is given by 
	\begin{equation}
		{U_I} =\frac{U^++U^-}{2}= \alpha \frac{{{G_0}}}{{\tau /{\tau _0} + 1}}\end{equation}
	
	Assuming $\Delta \tau \ll \tau$
	we arrive at the following expression for the relative spin-photo-EMF  (index $th$ stays for 'theory'): 
	\begin{equation}\xi_{th}  = \frac{U_S}{U_I} = -\gamma M{P_0}\frac{\tau }{{{\tau _0}}}\frac{1}{{\tau /{\tau _0} + 1}}
		\label{eq:eta}
	\end{equation}
	As it follows from (\ref{eq:eta}) the smaller tunneling time $\tau$ corresponding to the 
	thinner barrier results in a decrease of the spin-photo-EMF in full agreement with the experimental data.
	
	Let us now compare experimentally observed decrease of the spin-photo-EMF with decreasing the barrier thickness.
	The tunneling time is supposed to be exponentially dependent on the barrier thickness $d$:
	\[
	\tau^{-1}=T_0^{-1}\cdot e^{-2qd}, 
	\]
	where $q=\sqrt{2m_e W_e}/\hbar$, $m_e$ is the electron effective mass and $W_e$ is the potential barrier height for the electrons, $T_0$ is a  non-important prefactor.
	We take the parameters known for similar structures  as~\cite{OURPRB2015}  
	$W_e=45$ meV, $m_e=0.065 m_0$ to get 
	$q\approx 3\cdot10^6$ cm$^{-1}$.
	Assuming $\tau\ll \tau_0$,
	we can estimate the ratio between spin-photo-EMF for the two values of the barrier thickness ($5$, $10$~nm). This ratio is 
	\begin{equation}
		\frac{\xi_{th}(10 \text{ nm})}{{\xi_{th}(5 \text{ nm})}}\approx
		\frac{\tau(10 \text{ nm})}{\tau(5 \text{ nm})}\approx 20,
	\end{equation}
	which fully agrees with the experimental result. 
	
	\textbf{\textit{Optically detected FM proximity effect.}}
	Unlike spin-dependent part of the photo-EMF the degree of the spin polarization of the electrons in the QW would not decrease with decrease of the 
	tunneling time. 
	Indeed, let us consider linearly polarized or unpolarized  pumping. 
	The PL circular polarization degree due to spin polarization in the QW can be calculated as
	\begin{equation}{\left(\rho_c^\pi\right)_{th} } = -\frac{{n_ \uparrow ^ +
				+n_ \uparrow ^ -
				-
				n_ \downarrow ^ +-n_ \downarrow ^ - }}{{n_ \uparrow ^ + + n_ \uparrow ^ -  + n_ \downarrow ^ + + n_ \downarrow ^ - }}.\end{equation}
	With the above discussed assumptions we readily obtain 
	\begin{equation}\left(\rho_c^\pi\right)_{th} =   \frac{{\gamma M}}{2}\frac{1}{{1 + \tau /{\tau _0}}}.\end{equation}
	We note that a decrease of the barrier thickness would lead to a decrease of $\tau$ hence increase of the polarization. This appears to be in agreement with the experiment. 
	
	The experimentally measured
	intensity modulation corresponds to the difference in the total number of the carriers in the QW for left and right circular polarization. The corresponding quantity 
	to be calculated in the model is  
	\begin{equation}
		\eta_{th} = \frac{{n_\uparrow^+  + n_ \downarrow ^ +  - n_ \uparrow ^ -  - n_ \downarrow ^ - }}{{n_ \uparrow ^ +  + n_ \downarrow ^ +  + n_ \uparrow ^ -  + n_ \downarrow ^ - }}
	\end{equation}
	\begin{equation}\eta_{th} =   \frac{{P_0\gamma M}}{2}\frac{1}{{1 + \tau /{\tau _0}}}.\end{equation}
	
	The results obtained have a straightforward physical meaning. Circular polarization of the PL reflects the average spin of the electrons remaining in the QW. In its turn, the spin-photo-EMF is determined by the spin flux through the FM-SC interface, i.e. the difference in the electron fluxes with spins along and against the magnetization direction. Both quantities essentially depend on the ratio  of the mean tunnelling time $\tau$ to the radiative recombination time $\tau_0$. For a thick barrier $\tau/\tau_0\gg 1$, transport is inefficient and both the spin-photo-EMF and the circular polarization degree are small. In another limiting case of a thin barrier $\tau/\tau_0\ll 1$ spin-dependent transport dominates, polarization is maximum. However, in this case the spin-photo-EMF is again small, since tunneling becomes very efficient, and charge separation is almost complete regardless of the spin projection. The maximum value of the spin-photo-EMF can be expected when $\tau/\tau_0 \sim$1.

\textbf{Summary.} 
We have proposed a systematic approach to unambiguously isolate the tunneling spin-dependent transfer in FM-QW structures by combining optical and electrical measurements of the ferromagnetic proximity effect in the hybrid \mbox{GaMnAs/GaAs/InGaAs} heterostructure.
We emphasize that the two methods are  complementary and together represent a powerful technique to study the spin-dependent transport in  heterostructures similar to that under present study. In particular, for a  thin barrier, for which the PL signal under resonant excitation appeared too weak, the   photo-voltage signal allowed for the reliable measurements of the spin-dependent tunneling.
The experimentally observed dependence of the spin-photo-EMF on the tunnel barrier thickness is fully explained by a rate equations-based  theoretical model. We reveal and confirm the times mismatch phenomena very recently reported in a very different heterostructure. In our case this interplay between the tunneling rate and the recombination rate of the carriers is reflected in the    dependence of the measured quantities on the tunnel barrier thickness. At a thin tunnel barrier, electrons are efficiently captured with both spin orientations relative to the magnetic moment of the FM, which reduces the difference signal. On the contrary, the circular polarization of the PL from the QW increases, since the PL signal originates from a small number of electrons remaining in the QW, which are strongly polarized due to spin filtering by the ferromagnet.

Overall, the proposed systematic approach is directly applicable to other hybrid systems, i.e. based on 2D van der Waals materials or organic-inorganic semiconductor hybrids.
	
	\textbf{ASSOCIATED CONTENT}
	
	\textbf{Supporting Information.}
	S1 Methods: sample, polarized photoluminescence spectroscopy, photo-EMF and spin-photo-EMF measurement.
	S2 Role of magnetic circular dichroism.
	S3 Hybrid structure with $5$~nm barrier.
	
	\textbf{AUTHOR INFORMATION}
	
	Corresponding Author:
	
	rozhansky@gmail.com
	
	\textbf{ORCID}
	
	Igor V. Rozhansky: 0000-0001-9391-9304
	
	Ina V. Kalitukha: 0000-0003-2153-6667
	
	Grigorii S. Dimitriev: 0000-0002-6779-0997
	
	Olga S. Ken: 0000-0003-1541-863X
	
	Mikhail V. Dorokhin: 0000-0001-5238-0090
	
	Dmitri S. Arteev: 0000-0003-4447-8789
	
	Nikita S. Averkiev: 0000-0002-0772-7072
	
	Vladimir L. Korenev: 0000-0003-2839-2934
	
	\textbf{Notes}
	
	The authors declare no competing financial interest.
	
	\textbf{ACKNOWLEDGEMENTS}
	
	I.V.R, N.S.A acknowledge the financial support from the Russian Science Foundation, Project No.~22-12-00139 (theoretical calculations). The growth of magnetic structures was performed with the support of Russian Science Foundation, Project No.~21-79-20186. The experimental research was funded by the Russian Foundation for Basic Research (19-52-12034).
	
	\bibliography{references}
	
	\begin{tocentry}
		\includegraphics[width=8cm]{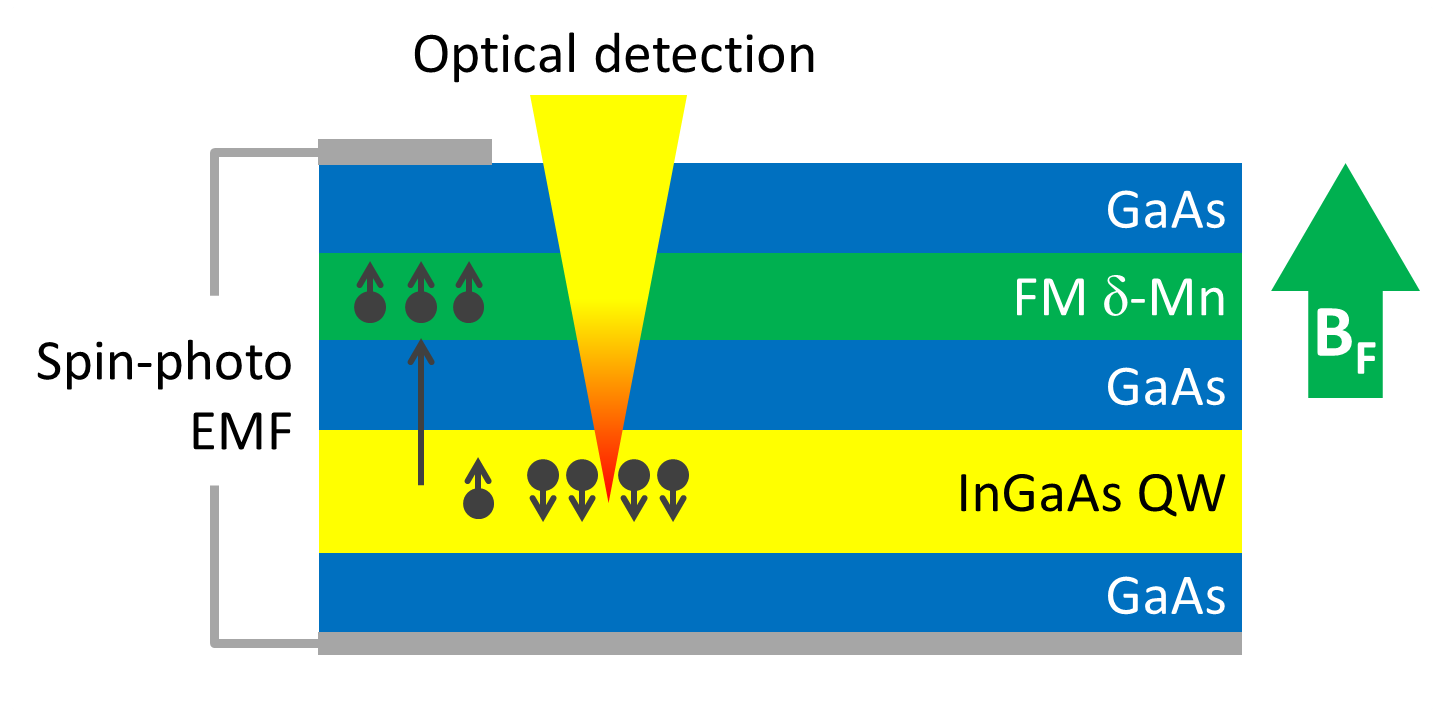}
	\end{tocentry}
	
\end{document}